\documentclass[a4paper,12pt]{article}
\usepackage{fullpage}
\usepackage{alltt}
\usepackage{url}
\title{Coq in a Hurry}
\author{Yves Bertot}
\date{October 2008}
\newcommand{\coqand}{/\char'134}
\newcommand{\coqor}{\char'134/}
\newcommand{\prodsym}{\mbox{forall }}
\newcommand{\coqnot}{\mbox{\~{}}}
\newcommand{\hide}[1]{}
\newcommand{\arrow}{\mbox{$\rightarrow$}}
\newcommand{\funarrow}{=>}
\begin{document}
\hfuzz=3pt
\maketitle
These notes provide a quick introduction to the Coq system and show how it
can be used to define logical concepts and functions and reason about them.
It is designed as a tutorial, so that readers can quickly start their own
experiments, learning only a few of the capabilities of the system.  A much
more comprehensive study is provided in \cite{coqart}, which also provides
an extensive collection of exercises to train on.
\section{Expressions and logical formulas}
The Coq system provides a language in which one handles formulas, verifies
that they are well-formed, and proves them.  Formulas may also contain functions
and limited forms of computations are provided for these functions.

The first thing you need to know is how you can check whether a formula is
well-formed.  The command is called {\tt Check}.  Here are a few examples,
which use a few of the basic objects and types of the system.  Commands
are always terminated by a period.
\begin{alltt}
Check True.
\textit{True : Prop}

Check False.
\textit{False : Prop}

Check 3.
\textit{3 : nat}

Check (3+4).
\textit{3 + 4 : nat}

Check (3=5).
\textit{3=5 : Prop}

Check (3,4).
\textit{(3,4) : nat * nat}

Check ((3=5)\coqand{}True).
\textit{3 = 5 \coqand{} True : Prop}

Check nat -> Prop.
\textit{nat -> Prop : Type}

Check (3 <= 6).
\textit{3 <= 6 : Prop}
\end{alltt}
The notation \(A {\tt :} B\) is uniformly used to indicate that the type of
the expression \(A\) is the expression \(B\).

Among these formulas, some can be read as propositions (they have type
{\tt Prop}), others may be read as numbers (they have type {\tt nat}), others
may be read as elements of more complex data structures.
You can also try to check badly formed formulas and in this case the Coq
system returns an informative error statement.

Complex formulas can be constructed by combining propositions with
logical connectives, or other expressions with addition,
multiplication, the pairing construct, and so on.  You can also
construct a new function by using the keyword {\tt fun}, which
replaces the \(\lambda\) symbol of lambda calculus and similar
theories.
\begin{alltt}
Check (fun x:nat => x = 3).
\textit{fun x : nat => x = 3 : nat -> Prop}

Check (forall x:nat, x < 3 \coqor{} (exists y:nat, x = y + 3)).
\textit{forall x : nat, x < 3 \coqor{} (exists y : nat, x = y + 3) : Prop}

Check (let f := fun x => (x * 3,x) in f 3).
\textit{let f := fun x : nat => (x * 3, x) in f 3 : nat * nat}
\end{alltt}
Please note that some notations are {\em overloaded}.  For instance,
the {\tt *} sign is used both to represent conventional multiplication on
numbers and the cartesian product on types.  One can find the function hidden
behind a notation by using the {\tt Locate} command.
\begin{alltt}
Locate "_ <= _".
\textit{Notation            Scope     }
\textit{"x <= y" := le x y   : nat_scope}
\textit{                      (default interpretation)}
\end{alltt}

The conditions for terms to be well-formed have two origins: first, the syntax
must be respected (parentheses, keywords, binary operators must have two
arguments); second, expressions must respect a type discipline.  The
{\tt Check} command not only checks that expressions are well-formed but it also
gives the type of expressions.  For instance we can use the {\tt Check} command
to verify progressively that some expressions are well-formed.
\begin{alltt}
Check True.
\textit{True : Prop}

Check False.
\textit{False : Prop}

Check and.
\textit{and : Prop -> Prop -> Prop}

Check (and True False).
\textit{True \coqand{} False : Prop}
\end{alltt}
In the last example, {\tt and} is a function that expects an argument of
type {\tt Prop} and returns a function of type {\tt Prop -> Prop}.  It can
therefore be applied to {\tt True}, which has the right type.  But the function
we obtain expects another argument of type {\tt Prop} and it can be applied
to the argument {\tt False}.  The notation
\begin{center}
 {\tt \(a\) -> \(b\) -> \(c\)}
\end{center}
actually stands for
\begin{center}
 {\tt \(a\) -> (\(b\) -> \(c\))}
\end{center}
 and the notation 
{\tt \(f\) \(a\) \(b\)} actually stands for {\tt (\(f\) \(a\)) \(b\)}.

The last example also shows that the notation {\tt /\char'134} is an infix
notation for the function {\tt and}.

Some constructs of the language have a notion of {\em bound} variable.
Among the examples we have already seen, the {\tt forall} and {\tt exists}
logical quantifiers and the {\tt fun} function constructor and the
{\tt let .. in} local declaration have bound variables.  When
constructs have a bound variable, this variable can be used
inside some part of the construct called {\em the scope}.  The type is
usually given explicitly, but it may also sometimes be left untold, and
the Coq system will infer it.

You can also require to evaluate an expression.  This actually means that
some symbolic computation is performed on this formula, and there are several
kind of strategies to perform this symbolic computation.  One strategy
is called {\tt compute}, here is an example of evaluation:
\begin{alltt}
Eval compute in 
  let f := fun x => (x * 3, x) in f 3.
\textit{= (9, 3) : nat * nat}
\end{alltt}
After executing this command, the function {\tt f} is not defined, it
was just defined temporarily inside the expression that was evaluated.
Here {\tt f} can be used only inside the expression {\tt f 3}.

At this point, you should be able to perform exercise 2.5 from \cite{coqart}
taking all numbers in the type {\tt nat}.
\begin{description}
\item[Exercise 2.5] Write a function that takes five arguments and returns
their sum.
\end{description}
\section{Defining new constants}
You can define a new constant by using the keyword {\tt Definition}.  Here
is an example:

\begin{alltt}
Definition example1 := fun x : nat => x*x+2*x+1.
\end{alltt}
An alternative, exactly equivalent, definition could be:
\begin{alltt}
Definition example1 (x : nat) := x*x+2*x+1.
\end{alltt}
After executing one of these commands, the function  can be used in other
commands:

\begin{alltt}
Check example1.
\textit{example1 : nat -> nat}

Eval compute in example 1.
\textit{= 4 : nat}
\end{alltt}
\section{Propositions and proofs}
The notation \(A {\tt :} B\) is actually used for several purposes in
the Coq system.  One of these purposes is to express that \(A\) is a
proof for the logical formula \(B\).  This habit is usually referred to under
the name {\em Curry-Howard Isomorphism}.  In accordance to this principle,
{\sf Prop} is actually a type whose elements are themselves types.  True
formulas are formulas that can be proved, in other words, types, in which
we know how to construct elements.  False formulas are types, in which
there is no way to construct elements.

To construct proofs of formulas, we simply look for existing proofs or use
functions to construct elements of a type from elements of other types.
\subsection{Finding existing proofs}
One can find already existing proofs of facts by using the {\tt Search} command.
Its argument should always be an identifier.
\begin{alltt}
Search True.
\textit{I : True}
Search le.
\textit{le_n : forall n : nat, n <= n}
\textit{le_S : forall n m : nat, le n m -> le n (S m)}
\end{alltt}
The theorem {\tt le\_S} uses a function {\tt S}, this function maps any
natural number to its successor.  Actually, the notation {\tt 3} is only
a notation for {\tt S (S (S O))}.

New theorems can be loaded from already proven packages using the
{\tt Require} command.  For example, for proofs in arithmetics, it is
useful to load the following packages:
\begin{alltt}
Require Import Arith Omega.
Search le.
\textit{between_le: forall (P : nat -> Prop) (k l : nat), 
   between P k l -> k <= l
exists_le_S:
  forall (Q : nat -> Prop) (k l : nat),
      exists_between Q k l -> S k <= l}
\dots
\textit{plus_le_reg_l: forall n m p : nat, p + n <= p + m -> n <= m
plus_le_compat_l: forall n m p : nat, n <= m -> p + n <= p + m
plus_le_compat_r: forall n m p : nat, n <= m -> n + p <= m + p
le_plus_l: forall n m : nat, n <= n + m}
\dots
\end{alltt}

The command {\tt SearchPattern} takes a pattern as argument, where some
of the arguments of a predicate can be replaced by incomplete expressions (we
use a special anonymous variable {\tt \_} to represent holes in incomplete
expressions).
\begin{alltt}
SearchPattern (_ + _ <= _ + _).
\textit{plus_le_compat_l: forall n m p: nat, n <= m -> p + n <= p + m
plus_le_compat_r: forall n m p: nat, n <= m -> n + p <= m + p
plus_le_compat: forall n m p q: nat, n <= m -> p <= q -> n + p <= m + q}
\end{alltt}
The command {\tt SearchRewrite} is similar, but it only looks for rewriting
theorems, that is, theorems where the proved predicate is an equality.  The
listed theorems are those for which the pattern fits one of the
members of the equality in the conclusion.
\begin{alltt}
SearchRewrite (_ + (_ - _)).
\textit{le_plus_minus: forall n m : nat, n <= m -> m = n + (m - n)
le_plus_minus_r: forall n m : nat, n <= m -> n + (m - n) = m}
\end{alltt}

\subsection{Constructing new proofs}
A theorem that proves an implication \(A\Rightarrow B\) actually
is an expression whose
type is a function type {\tt \(A\) -> \(B\)} (in other words, an arrow type).
Such a function (let's call it \(f\))
 can be applied to a term \(x\) of type \(A\).  From the logical
point of view, when applying the function to the argument to form \(f~x\),
we actually deduce that \(B\) holds from the
proofs that \(A\) and \hbox{\tt \(A\) -> \(B\)} hold.
This corresponds to what is usually known as {\em modus ponens}.

A theorem that proves a universal quantification is also a function.
It can also be applied to an argument of the right
type.  For instance, if the theorem \(t\) is a proof of 
\(\forall x:A,~P~x\) and \(a\) has type \(A\), then we can apply \(t\) to
\(A\).  The resulting expression, \(t~a\), has the type \(P~a\).
From the logical point of view, this corresponds to producing a new
proof for the statement where the universal quantification is instantiated.
Here are a few examples, where theorems are instantiated and a modus ponens
inference is performed:
\begin{alltt}
Check le_n.
\textit{le_n : forall n:nat n <= n}

Check (le_n 0).
\textit{le_n 0 : 0 <= 0}

Check (le_S 0 0).
\textit{le_S 0 1 : 0 <= 0 -> 0 <= 1}

Check (le_S 0 0 (le_n 0)).
\textit{le_S 0 0 (le_n O) : 0 <= 1}
\end{alltt}
New theorems could be constructed this way by combining existing theorems and
using the {\tt Definition} keyword to associate these expressions to
constants.  But this approach is seldom used.  The alternative approach
is known as {\em goal directed proof}, with the following type of
scenario:
\begin{enumerate}
\item the user enters a statement that he wants to prove, using the command
{\tt Theorem} or {\tt Lemma},
\item the Coq system displays the formula as a formula to be proved, possibly
giving a context of local facts that can be used for this proof (the context
is displayed above a horizontal line, the goal is displayed under the
horizontal line),
\item the user enters a command to decompose the goal into simpler ones,
\item the Coq system displays a list of formulas that still need to be proved,
\item back to step 3.
\end{enumerate}
Some of the commands sent at step 3 actually decrease the number of goals.
When there are no more goals the proof is complete, it needs to be saved, this
is performed when the user sends the command {\tt Qed.}  The commands that
are especially designed to decompose goals into collections of simpler goals
are called {\em tactics}.

Here is an example:
\begin{alltt}
Theorem example2 : forall a b:Prop, a \coqand{} b -> b \coqand{} a.
\textit{1 subgoal
  
  ============================
   forall a b : Prop, a \coqand{} b -> b \coqand{} a
}
Proof.
 intros a b H.
\textit{1 subgoal
  
  a : Prop
  b : Prop
  H : a \coqand{} b
  ============================
   b \coqand{} a}

split.
\textit{2 subgoals}
\dots
\textit{  H : a \coqand{} b
  ============================
   b

subgoal 2 is:
 a
}
elim H; intros H0 H1.
\dots
\textit{  H0 : a
  H1 : b
  ============================
   b
}
exact H1.
\textit{1 subgoal}
\dots
\textit{  H : a \coqand{} b
  ============================
   a
}
intuition.
\textit{Proof completed.}

Qed.
\textit{intros a b H.
split.
 elim H; intros H0 H1.
   exact H1.
 intuition.
example2 is defined}
\end{alltt}
This proof uses several steps to decompose the logical processing, but
a quicker dialog would simply rely on the automatic tactic {\tt intuition}
directly from the start.  There is an important collection of tactics
in the Coq system, each of which is adapted to a shape of goal.  For
instance, the tactic {\tt elim H} was adapted because the hypothesis
{\tt H} was a proof of a conjunction of two propositions.  The effect
of the tactic was to add two implications in front of the goal, with
two premises asserting one of the propositions in the conjunction.  It
is worthwhile remembering a collection of tactics for the basic
logical connectives.  We list these tactics in the following table,
inspired from the table in \cite{coqart} (p. 130).
\begin{center}
\begin{tabular}{|c|c|c|c|}
\hline
&\(\Rightarrow\)&\(\forall\)&\(\wedge\)\\
\hline
Hypothesis {\tt H}&{\tt apply H}&{\tt apply H}&{\tt elim H }\\
&&&{\tt case H}\\
&&&{\tt destruct H as [H1 H2]}\\
\hline
goal&{\tt intros {\tt H}} &{\tt intros {\tt H}} &{\tt split}\\
\hline
&\(\neg\)&\(\exists\)&\(\vee\)\\
\hline
Hypothesis {\tt H}&{\tt elim H}&{\tt elim H}&{\tt elim H}\\
&{\tt case H}&{\tt case H}&{\tt case H}\\
&&{\tt destruct H as [x H1]}&{\tt destruct H as [H1 | H2]}\\
\hline
goal&{\tt intros H}&{\tt exists} {\it v}&{\tt left}
or\\
&&&{\tt right}\\
\hline
&&\(=\)&\\
\hline
Hypothesis {\tt H}&&{\tt rewrite H}&\\
&&{\tt rewrite <- H}&\\
\hline
goal&&{\tt reflexivity}&\\
&&{\tt ring}&\\
\hline
\end{tabular}
\end{center}
\label{table}
When using the tactic {\tt elim} or {\tt case}, this usually creates new
facts that
are placed in the result goal as premises of newly created implications.
These premises must then be introduced in the context using the {\tt intros}
tactic.  A quicker tactic does the two operations at once, this tactic
is called {\tt destruct}.

A common approach to proving difficult propositions is to assert
intermediary steps using a tactic called {\tt assert}.  Given
an argument of the form {\tt (H : \(P\))}, this tactic yields two goals,
where the first one is to prove \(P\), while the second one is to prove
the same statement as before, but in a context where an hypothesis
named {\tt H} and with the statement \(P\) is added.

Some automatic tactics are also provided for a variety of purposes,
{\tt intuition} is often useful to prove facts that are tautologies in
first-order intuitionistic logic (try it whenever the proof only involves
manipulations of forall quantification, conjunction, disjunction, and
 negation); {\tt auto} is an extensible tactic that tries to apply a collection
of theorems that were provided beforehand by the user, {\tt eauto} is like
{\tt auto}, it is more powerful but also more time-consuming,  {\tt ring}
mostly does proofs of equality for expressions containing
addition and multiplication, {\tt omega} proves systems of linear inequations.
To use {\tt omega} you first need to require that Coq loads it into memory.
An inequation
is linear when it is of the form \(A\leq B\) or \(A<B\), and \(A\) and \(B\)
are sums of terms of the form \(n*x\), where \(x\) is a variable and \(n\)
is a numeric constant, 1, 3, -15, etcetera).  The tactic {\tt omega} can
also be used for problems, which are {\em instances} of systems of linear 
inequations.  Here is an example, where {\tt f x} is
not a linear term, but the whole system of inequations is the instance
of a linear system.
\begin{alltt}
Require Import Omega.

Lemma omega_example : 
  forall f x y, 0 < x -> 0 < f x -> 3 * f x <= 2 * y -> f x <= y.
intros; omega.
Qed.
\end{alltt}

One of the difficult points for newcomers is that the Coq system also provides
a type {\tt bool} with two elements called {\tt true} and {\tt false}, but
traditionally does not use these boolean values to represent the difference
between provable and unprovable propositions.  The type {\tt bool}  is just
a two-element type that may be used to model the boolean types that one usually
finds in programming languages and the value attached to its two elements
is purely conventional.  In general, a proposition is represented as a type,
and the fact that this proposition holds is represented by the existence
of an element in this type.  In practice, the proposition (which is a type)
cannot be used as a boolean value, for instance in a conditional statement,
and a boolean value, which is atomic and not a type, cannot contain a proof.
In some cases, the two points of view can be reconciled, especially in
{\em proof by reflexion}, but we will not cover this topic in this tutorial.

At this point, you should be able to perform the following two
exercises.
\begin{description}
\item[Exercise 5.6]  (from \cite{coqart}) Prove the following theorems:
\begin{alltt}
\hide{Check (}{\prodsym}A B C:Prop, A\coqand{}(B\coqand{}C)\arrow{}(A\coqand{}B)\coqand{}C
\hide{). Check (}{\prodsym}A B C D: Prop,(A\arrow{}B)\coqand{}(C\arrow{}D)\coqand{}A\coqand{}C \arrow{} B\coqand{}D
\hide{). Check (}{\prodsym}A: Prop, \coqnot(A\coqand{}\coqnot{}A)
\hide{). Check (}{\prodsym}A B C: Prop, A\coqor{}(B\coqor{}C)\arrow{}(A\coqor{}B)\coqor{}C
\hide{). Check (}{\prodsym}A: Prop, \coqnot{}\coqnot{}(A\coqor{}\coqnot{}A)
\hide{). Check (}{\prodsym}A B: Prop, (A\coqor{}B)\coqand{}\coqnot{}A \arrow{} B\hide{).}
\end{alltt}
Two benefits can be taken from this exercise.  In a first step you
should try using only the basic tactics given in the table
page~\pageref{table}.  In a second step, you can verify which of these
statements are directly solved by the tactic {\tt intuition}.
\item[Universal quantification] Prove
\begin{alltt}
{\prodsym}A:Set,{\prodsym}P Q:A\arrow{}Prop,
    ({\prodsym}x, P x)\coqor{}({\prodsym}y, Q y)\arrow{}{\prodsym}x, P x\coqor{}Q x.
\end{alltt}
\end{description}
\section{Inductive types}
Inductive types could also be called algebraic types or initial algebras.
They are defined by providing the type name, its type, and a collection
of constructors.  They are often used to model data-structures.
Inductive types can be parametrized and dependent.
We will mostly use parametrization to represent
polymorphism and dependence to represent logical properties.

\subsection{Defining inductive types}
Here is an example of an inductive type definition:
\begin{alltt}
Inductive bin : Set :=
  L : bin
| N : bin -> bin -> bin.
\end{alltt}
This defines a new type {\tt bin}, whose type is {\tt Set}, and provides
two ways to construct elements of this type: {\tt L} (a constant) and
{\tt N} (a function taking two arguments).  The Coq system automatically
associates a theorem to this inductive type.  This theorem makes it possible
to reason by induction on elements of this type:
\begin{alltt}
Check bin_ind.
\textit{bin_ind
     : forall P : bin -> Prop,
       P L ->
       (forall b : bin,
         P b -> forall b0 : bin, P b0 -> P (N b b0)) ->
       forall b : bin, P b}
\end{alltt}
The induction theorem associated to an inductive type is always named
{\tt {\it name}\_ind}.

\subsection{Pattern matching}
Elements of inductive types can be processed using functions that perform
some pattern-matching.  For instance, we can write a function that
returns the boolean value {\tt false} when its argument is {\tt N L L}
and returns {\tt true} otherwise.

\begin{alltt}
Definition example3 (t : bin): bool :=
  match t with N L L => false | _ => true end.
\end{alltt}

\subsection{Recursive function definition}
There are an infinity of different trees in the type {\tt bin}.  To write
interesting functions with arguments from this type, we need more than just
pattern matching.  The Coq system provides recursive programming.  The shape
of recursive function definitions is as follows:
\begin{alltt}
Fixpoint flatten_aux (t1 t2:bin) \{struct t1\} : bin :=
  match t1 with
   L => N L t2   |   N t'1 t'2 => flatten_aux t'1 (flatten_aux t'2 t2)
  end.

Fixpoint flatten (t:bin) : bin :=
  match t with
    L => L | N t1 t2 => flatten_aux t1 (flatten t2)
  end.

Fixpoint size (t:bin) : nat :=
  match t with
    L => 1 | N t1 t2 => 1 + size t1 + size t2
  end.
\end{alltt}
There are constraints in the definition of recursive definitions.  First,
if the function has more than one argument, we must declare one of
these arguments as the {\em principal} or {\em structural} argument
(we did this declaration for the function {\tt flatten\_aux}).
If there is only one argument that belongs to an inductive type, then this
argument is the principal argument
by default.  Second, every recursive call must be performed so that the
principal argument of the recursive call is a sub-term, obtained by 
pattern-matching, of the initial principal argument.  This condition is
satisfied in all the examples above.

In computation, recursive function are easy to simplify when their structural
argument has the form of a constructor, as appears in the following case:
\begin{alltt}
Eval compute in flatten_aux (N L L) L.
\textit{ = N L (N L L) : bin}
\end{alltt}
\subsection{Proof by cases}
Now, that we have defined functions on our inductive type, we can prove
properties of these functions.  Here is a first example, where we perform
a few case analyses on the elements of the type {\tt bin}.

\begin{alltt}
Theorem example3_size :
   forall t, example3 t = false -> size t = 3.
Proof.
intros t; destruct t.
\textit{2 subgoals
  
  ============================
   example3 L = false -> size L = 3

subgoal 2 is:
 example3 (N t1 t2) = false -> size (N t1 t2) = 3}
\end{alltt}
The tactic {\tt destruct t} actually observes the various possible cases
for {\tt t} according to the inductive type definition.  The term {\tt t}
can only be either obtained by {\tt L}, or obtained by {\tt N} applied to
two other trees {\tt t1} and {\tt t2}.  This is the reason why there
are two subgoals.

We know the value that {\tt example3} and {\tt size} should take for the
tree {\tt L}.  We can direct the Coq system to compute it:
\begin{alltt}
simpl.
\textit{2 subgoals
  
  ============================
   true = false -> 1 = 3}
\end{alltt}
After computation, we discover that assuming that the tree is {\tt L} and
that the value of {\tt example3} for this tree is {\tt false} leads to
an inconsistency.  We can use the following tactics to exploit this kind
of inconsistency:
\begin{alltt}
intros H.
\textit{  H : true = false
  ============================
   1 = 3}
discriminate H.
\textit{1 subgoal}
\dots
\textit{  ============================
   example3 (N t1 t2) = false -> size (N t1 t2) = 3}
\end{alltt}
The answer shows that the first goal was solved.  The tactic 
{\tt discriminate H} can be used whenever the hypothesis {\tt H} is an
assumption that asserts that two different constructors of an inductive
type return equal values.  Such an assumption is inconsistent and the tactic
exploits directly this inconsistency to express that the case described
in this goal can never happen.  This tactic expresses a basic property
of inductive types in the sort {\tt Set} or {\tt Type}: constructors have
distinct ranges.  Another important property of constructors of inductive
types in the sort {\tt Set} or {\tt Type} is that they are injective.  The
tactic to exploit this fact called {\tt injection} (for more details about
{\tt discriminate} and {\tt injection} please refer to \cite{coqart} or the
Coq reference manual \cite{coqmanual}).

For the second goal
we still must do a case analysis on the values of {\tt t1} and {\tt t2},
we do not detail the proof but it can be completed with the following
sequence of tactics.
\begin{alltt}
destruct t1.
destruct t2.
\textit{3 subgoals
  
  ============================
   example3 (N L L) = false -> size (N L L) = 3

subgoal 2 is:
 example3 (N L (N t2_1 t2_2)) = false -> 
 size (N L (N t2_1 t2_2)) = 3
subgoal 3 is:
 example3 (N (N t1_1 t1_2) t2) = false ->
  size (N (N t1_1 t1_2) t2) = 3}
\end{alltt}
For the first goal, we know that both functions will compute as we
stated by the equality's right hand side.  We can solve this goal
easily, for instance with {\tt auto}.  The last two goals are solved
in the same manner as the very first one, because {\tt example3}
cannot possibly have the value {\tt false} for the arguments that are
given in these goals.
\begin{alltt}
auto.
intros H; discriminate H.
intros H; discriminate H.
Qed.
\end{alltt}
To perform this proof, we have simply observed 5 cases.  For general recursive
functions, just observing a finite number of cases is not sufficient.  We
need to perform proofs by induction.

At this point, you can perform a second part of our exercise on universal
quantification.
\begin{description}
\item[Exercise on universal quantification (follow-up)]  Prove the following
theorem
\begin{alltt}
\coqnot({\prodsym}A:Set,{\prodsym}P Q:A\arrow{}Prop,
   ({\prodsym}x, P x\coqor{}Q x)\arrow{}({\prodsym}x, P x)\coqor{}({\prodsym}y, Q y).
\end{alltt}
\end{description}
Hint: think about a counter-example with the type
{\tt bool} and its two elements {\tt true} and {\tt false}, which can be
proved different using {\tt discriminate}.

\subsection{Proof by induction}
The most general kind of proof that one can perform on inductive types is
proof by induction.

When we prove a property of the elements of an inductive type using a proof by
induction, we actually consider a case for each constructor, as we did
for proofs by cases.  However there is a twist: when we consider a constructor
that has arguments of the inductive type, we can assume that the property
we want to establish holds for each of these arguments.

When we do goal directed proof, the induction principle is invoked by the
{\tt elim} tactic.  To illustrate this tactic, we will prove a simple
fact about the {\tt flatten\_aux} and {\tt size} functions.
\begin{alltt}
Theorem flatten_aux_size : 
 forall t1 t2, size (flatten_aux t1 t2) = size t1 + size t2 + 1.
Proof.
 intros t1; elim t1.
\textit{  ============================
   forall t2 : bin, size (flatten_aux L t2) = size L + size t2 + 1

subgoal 2 is:
 forall b : bin,
 (forall t2 : bin, size (flatten_aux b t2) =
                   size b + size t2 + 1) ->
 forall b0 : bin,
 (forall t2 : bin, size (flatten_aux b0 t2) =
                   size b0 + size t2 + 1) ->
 forall t2 : bin,
 size (flatten_aux (N b b0) t2) =
 size (N b b0) + size t2 + 1
}
\end{alltt}
There are two subgoals, the first goal requires that we prove the property
when the first argument of {\tt flatten\_aux} is {\tt L}, the second one
requires that we prove the property when the argument is {\tt N b b0}, under
the assumption that it is already true for {\tt b} and {\tt b0}.
The proof progresses easily, using the definitions of the two functions,
which are expanded when the Coq system executes the {\tt simpl} tactic.
We then obtain expressions that can be solved using the {\tt ring} tactic.

\begin{alltt}
intros t2.
simpl.
\dots
\textit{  ============================
   S (S (size t2)) = S (size t2 + 1)
}
ring.
intros b IHb b0 IHb0 t2.
\dots
\textit{  IHb : forall t2 : bin, size (flatten_aux b t2) =
                                 size b + size t2 + 1
  b0 : bin
  IHb0 : forall t2 : bin, size (flatten_aux b0 t2) = 
                          size b0 + size t2 + 1
  t2 : bin
  ============================
   size (flatten_aux (N b b0) t2) = size (N b b0) + size t2 + 1
} 
simpl.
\dots
\textit{  ============================
   size (flatten_aux b (flatten_aux b0 t2)) =
   S (size b + size b0 + size t2 + 1)
}
rewrite IHb.
\dots
\textit{  ============================
   size b + size (flatten_aux b0 t2) + 1 = 
   S (size b + size b0 + size t2 + 1)
}
rewrite IHb0.
\dots
\textit{  ============================
   size b + (size b0 + size t2 + 1) + 1 = 
   S (size b + size b0 + size t2 + 1)
}
ring.
\textit{Proof completed.}
Qed.
\end{alltt}
At this point, you should be able to perform your own proof by induction.
\begin{description}
\item[Exercise on {\tt flatten} and {\tt size}] Prove
\begin{alltt}
Theorem flatten_size : forall t, size (flatten t) = size t.
\end{alltt}
\end{description}
\subsection{Numbers in the Coq system}
In the Coq system, most usual data-types are represented as inductive types
and packages provide a variety of properties, functions, and theorems around
these data-types.  The package named {\tt Arith} contains a host of theorems
about natural numbers (numbers from 0 to infinity), which are described
as an inductive type with {\tt O} (representing 0) and {\tt S} as constructors.
It also provides addition, multiplication, subtraction (with the special
behavior that {\tt x - y} is 0 when {\tt x} is smaller than {\tt y}).
This package also provides a tactic {\tt ring}, which solves equalities
between expressions modulo associativity and commutativity of addition and
multiplication and distributivity.  For natural numbers, {\tt ring} does
not handle subtraction well, because of its special behavior.

The package named {\tt ZArith} provides two inductive
data-types to represent integers.  The first inductive type, named
 {\tt positive}, follows a binary representation to model the positive integers
(from 1 to infinity) and the type {\tt Z} is described as a type with three
constructors, one for positive numbers, one for negative numbers, and one
for 0.  The package also provides orders and basic operations: addition,
subtraction, multiplication, division, square root.  The tactic {\tt ring}
also works for integers (this time, subtraction is well supported)  The tactic 
{\tt omega} works equally well to solve problems in Presburger arithmetic
for both natural numbers of type {\tt nat} and integers of type {\tt Z}.

There is also a package {\tt QArith} for rational numbers and a package
{\tt Reals} for real numbers.  While the former is 
quite rudimentary, the latter is quite extensive, up to derivation and
trigonometric functions, for instance.  For these packages, a tactic
{\tt field} helps solving equalities between fractional expressions, and
a tactic {\tt fourier} helps solving systems of linear inequations.

\subsection{Data-structures}
The two-element boolean type is an inductive type in the
Coq system, {\tt true} and {\tt false} are its constructors.  The
induction principle naturally expresses that this type only has two elements,
because it suffices that a property is satisfied by {\tt true} and {\tt false}
to ensure that it is satisfied by all elements of {\tt bool}.  On the other
hand, it is easy to prove that {\tt true} and {\tt false} are distinct.

Most ways to structure data together are also provided using inductive data
structures.  The pairing construct actually is an inductive type, and
{\tt elim} can be used to reason about a pair in the same manner as it
can be used to reason on a natural number.

A commonly used data-type is the type of lists.  This type is
polymorphic, in the sense that the same inductive type can be used for
lists of natural numbers, lists of boolean values, or lists of other
lists.  This type is not provided by default in the Coq system, it is
necessary to load the package {\tt List} using the {\tt Require}
command to have access to it.  The usual {\tt cons} function is given
in Coq as a three argument function, where the first argument is the
type of elements: it is the type of the second argument and the third
argument should be a list of elements of this type, too.  The empty
list is represented by a function {\tt nil}.  This function also takes
an argument, which should be a type.  However, the Coq system also
provides a notion of implicit arguments, so that the type arguments
are almost never written and the Coq system infers them from the
context or the other arguments.  For instance, here is how we
construct a list of natural numbers.
\begin{alltt}
Require Import List.

Check (cons 3 (cons 2 (cons 1 nil))).
\textit{3 :: 2 :: 1 :: nil : list nat}
\end{alltt}
This example also shows that the notation {\tt ::} is used to represent the
cons function in an infix fashion.  This tradition will be very comfortable
to programmers accustomed to languages in the ML family, but Haskell addicts
should beware that the conventions, between type information and cons, are
inverse to the conventions in Haskell.

The {\tt List} package also provides a list concatenation function named
{\tt app}, with {\tt ++} as infix notation, and a few theorems about this
function.
\section{Inductive properties}
Inductive types can be dependent and when they are, they can be used to
express logical properties.  When defining inductive types like {\tt nat},
{\tt Z} or {\tt bool} we declare that this constant is a type.  When defining
a dependent type, we actually introduce a new constant which is declared to
be a function from some input type to a type of types.  Here is an example:

\begin{alltt}
Inductive even : nat -> Prop :=
  even0 : even 0
| evenS : forall x:nat, even x -> even (S (S x)).
\end{alltt}
Thus, {\tt even} itself is not a type, it is {\tt even x}, whenever {\tt x}
is an integer that is a type.  In other words, we actually defined a family
of types.  In this family, not all members contain elements.  For instance,
we know that the type {\tt even 0} contains an element, this element
is {\tt even0}, and we know that {\tt even 2} contains an element:
{\tt evenS 0 even0}.  What about {\tt even 1}?  If {\tt even 1} contains
an element, this element cannot be obtained using {\tt even0} because 
\(0\neq 1\), it cannot be obtained using {\tt evenS} because \(1\neq 2+x\)
for every \(x\) such that \(0 \leq x\).  Pushing our study of {\tt even}
further we could see that this type, seen as a property, is provable if and
only if its argument is even, in the common mathematical sense.

Like other inductive types, inductive properties are equipped with an
induction principle, which we can use to perform proofs.  The inductive
principle for {\tt even} has the following shape.
\begin{alltt}
Check even_ind.
\textit{even_ind : forall P : nat -> Prop,
       P 0 ->
       (forall x : nat, even x -> P x -> P (S (S x))) ->
       forall n : nat, even n -> P n}
\end{alltt}
This principle intuitively expresses that {\tt even} is the smallest property
satisfying the two constructors: it implies every other property that also
satisfies them.  A proof using this induction principle will work on a goal
where we know that {\tt even y} holds for some {\tt y} and actually decomposes
into a proof of two cases, one corresponding to the case where {\tt even y}
was obtained using {\tt even0} and one corresponding to the case where
{\tt even y} was obtained using {\tt evenS}, applied to some {\tt x} such
that {\tt y=S (S x)} and some proof of {\tt even y}.  In the second case,
we again have the opportunity to use an induction hypothesis about this
{\tt y}.

When a variable {\tt x} satisfies an inductive property, it is often more
efficient to prove properties about this variable using an induction
on the inductive property than an induction on the variable itself.  The
following proof is an example:
\begin{alltt}
Theorem even_mult : forall x, even x -> exists y, x = 2*y.
Proof.
intros x H; elim H.
\textit{2 subgoals
  
  x : nat
  H : even x
  ============================
   exists y : nat, 0 = 2 * y

subgoal 2 is:
 forall x0 : nat,
 even x0 -> (exists y : nat, x0 = 2 * y) -> 
 exists y : nat, S (S x0) = 2 * y
}
exists 0; ring.
intros x0 Hevenx0 IHx.
\dots
\textit{  IHx : exists y : nat, x0 = 2 * y
  ============================
   exists y : nat, S (S x0) = 2 * y
}
\end{alltt}
In the last goal, {\tt IHx} is the induction hypothesis.  It says that
if {\tt x0} is the predecessor's predecessor of {\tt x} then we
already know that there exists a value {\tt y} that is its half.  We
can use this value to provide the half of {\tt S (S x0)}.  Here are
the tactics that complete the proof.
\begin{alltt}
destruct IHx as [y Heq]; rewrite Heq.
exists (S y); ring.
Qed.
\end{alltt}
In this example, we used a variant of the {\tt destruct} tactic that
makes it possible to choose the name of the elements that {\tt destruct}
creates and introduces in the context.

If we wanted to prove the same property using a direct induction on the
natural number that is even, we would have a problem because the predecessor
of an even number, for which direct induction provides an induction hypothesis,
is not even.  The proof is not impossible but slightly more complex:
\begin{alltt}
Theorem even_mult' : forall x, even x -> exists y, x = 2* y.
Proof.
intros x.
assert (lemma: (even x -> exists y, x=2*y)\coqand
        (even (S x) -> exists y, S x=2*y)).
elim x.
split.
exists 0; ring.
intros Heven1; inversion Heven1.
intros x0 IHx0; destruct IHx0 as [IHx0 IHSx0].
split.
exact IHSx0.
intros HevenSSx0.
assert (Hevenx0 : even x0).
inversion HevenSSx0; assumption.
destruct (IHx0 Hevenx0) as [y Heq].
rewrite Heq; exists (S y); ring.
intuition.
Qed.
\end{alltt}
This script mostly uses tactics that we have already introduced, except the
{\tt inversion} tactic.  Given an assumption {\tt H} that relies on a
dependent inductive type, most frequently an inductive proposition, the
tactic {\tt inversion} analyses all the constructors of the inductive, discards
the ones that could not have been applied, and when some constructors could
have applied it creates a new goal where the premises of this constructor
are added in the context.  For instance, this tactic is perfectly suited
to prove that 1 is not even:
\begin{alltt}
Theorem not_even_1 : ~even 1.
Proof.
intros even1.
\dots
\textit{  even1 : even 1
  ============================
   False}

inversion even1.
Qed.
\end{alltt}
This example also shows that the negation of a fact actually is represented
by a function that says ``this fact implies {\tt False}''.

Inductive properties can be used to express very complex notions.  For
instance, the semantics of a programming language can be defined as an
inductive definition, using dozens of constructors, each one describing
a kind of computation elementary step.  Proofs by induction with respect
to this inductive definition correspond to what Wynskel calls
{\em rule induction}  in his introductory book on programming language
semantics \cite{winskel93}.
\section{Exercises}
Most of the exercises proposed here are taken from \cite{coqart}.  We
also repeat exercises that were already given in the text.
\begin{description}
\item[Exercise 2.5] Write a function that takes five arguments and returns
their sum.
\item[Exercise 5.6]  Prove the following theorems:
\begin{alltt}
\hide{Check (}{\prodsym}A B C:Prop, A\coqand{}(B\coqand{}C)\arrow{}(A\coqand{}B)\coqand{}C
\hide{). Check (}{\prodsym}A B C D: Prop,(A\arrow{}B)\coqand{}(C\arrow{}D)\coqand{}A\coqand{}C \arrow{} B\coqand{}D
\hide{). Check (}{\prodsym}A: Prop, \coqnot(A\coqand{}\coqnot{}A)
\hide{). Check (}{\prodsym}A B C: Prop, A\coqor{}(B\coqor{}C)\arrow{}(A\coqor{}B)\coqor{}C
\hide{). Check (}{\prodsym}A: Prop, \coqnot{}\coqnot{}(A\coqor{}\coqnot{}A)
\hide{). Check (}{\prodsym}A B: Prop, (A\coqor{}B)\coqand{}\coqnot{}A \arrow{} B\hide{).}
\end{alltt}
Two benefits can be taken from this exercise.  In a first step you
should try using only the basic tactics given in the table
page~\pageref{table}.  In a second step, you can verify which of these
statements are directly solved by the tactic {\tt intuition}.
\item[Universal quantification] Prove
\begin{alltt}
{\prodsym}A:Set,{\prodsym}P Q:A\arrow{}Prop,
    ({\prodsym}x, P x)\coqor{}({\prodsym}y, Q y)\arrow{}{\prodsym}x, P x\coqor{}Q x.
\coqnot({\prodsym}A:Set,{\prodsym}P Q:A\arrow{}Prop,
   ({\prodsym}x, P x\coqor{}Q x)\arrow{}({\prodsym}x, P x)\coqor{}({\prodsym}y, Q y).
\end{alltt}
Hint: for the second exercise, think about a counter-example with the type
{\tt bool} and its two elements {\tt true} and {\tt false}, which can be
proved different, for instance.
\item[Exercise 6.32]The sum of the first \(n\) natural numbers is defined with the
following function:
\begin{alltt}
Fixpoint sum_n (n:nat) : nat :=
  match n with  0 \funarrow{} 0 | S p \funarrow{} S p + sum_n p end.
\end{alltt}
Prove the following statement:
\begin{alltt}
\hide{Check (}{\prodsym}n:nat, 2 * sum_n n = n*n + n\hide{).}
\end{alltt}
\item[Sum of powers] Define a recursive function {\tt nat\_power}.  Find in
the Coq documentation\footnote{\url{http://coq.inria.fr/}} how to attach the
notation {\tt x\^{}y} to this function.
Then define a summation function that makes it possible to compute
\[\sum_{k=0}^n f(k)\]
Prove the following theorem:
\[\forall x~ n:nat, 1 \leq x \rightarrow (x-1)\sum_{k=0}^n x^k=x^{n+1}-1.\]
To perform this exercise, we suggest you first prove an equivalent statement
where no subtraction occurs (by induction over \(n\), using {\tt simpl}
{\tt ring}, and {\tt simpl} to perform the computational steps).  You can
then switch to the statement with subtractions using the known theorems
{\tt mult\_reg\_l} and {\tt le\_plus\_minus}.
\item[Square root of 2]  If \(p^2=2q^2\), then \((2q-p)^2=2(p-q)^2\), now
if \(p\) is the least positive integer such that there exists a positive
integer \(q\) such that
\(p/q=\sqrt{2}\), then \(p>2q-p>0\), and \((2q-p)/(p-q)=\sqrt{2}\).  This is a
contradiction and a proof that \(\sqrt{2}\) is not rational.  Use
Coq to verify a formal proof along these lines (use {\tt Z} as the type
of numbers).
\item[Sum of powers revisited]
Try to redo this exercise with other number types, like {\tt Z}, {\tt R},
where powers and summations may already be defined and the condition
\(1\leq x\) can usually be replaced by another condition.
\end{description}
\section{Solutions}
The solutions to the numbered exercises are available from the Internet
(on the site associated to the reference \cite{coqart}).
The short proof that \(\sqrt{2}\) is not rational is also available on
the Internet\footnote{\url{http://cocorico.cs.ru.nl/coqwiki/SquareRootTwo}}

\subsection{Universal quantification}
Here are the solutions to the exercises on universal quantification.
{\small
\begin{alltt}
Theorem ex1 :
  forall A:Set, forall P Q:A->Prop,
  (forall x, P x) \coqor{} (forall y, Q y) -> forall x, P x \coqor{} Q x.
Proof.
 intros A P Q H.
 elim H.
 intros H1; left; apply H1.
 intros H2; right; apply H2.
Qed.

Theorem ex2 :
  ~(forall A:Set, forall P Q:A->Prop,
    (forall x, P x \coqor{} Q x) -> (forall x, P x) \coqor{} (forall y, Q y)).
Proof.
  intros H; elim (H bool (fun x:bool => x = true)
                    (fun x:bool => x = false)).
  intros H1; assert (H2:false = true).
    apply H1.
  discriminate H2.
  intros H1; assert (H2:true = false).
    apply H1.
  discriminate H2.
  intros x; case x.
  left; reflexivity.
  right; reflexivity.
Qed.
\end{alltt}
}
\subsection{flatten}
Here is the solution to the exercise on {\tt flatten} and {\tt size}
(this re-uses the lemma\\ {\tt flatten\_aux\_size} proved in these notes).
{\small
\begin{alltt}
Theorem flatten_size :  forall t, size(flatten t) = size t.
Proof.
intros t; elim t.
simpl. reflexivity.
intros t1 IH1 t2 IH2; simpl. rewrite flatten_aux_size. rewrite IH2. ring.
Qed.
\end{alltt}
}
\subsection{Sum of powers}
Here is a condensed solution to the exercise on the sum of powers:
{\small
\begin{alltt}
Require Import Arith Omega.

Fixpoint nat_power (n m:nat) \{struct m\} : nat :=
 match m with 0 => 1 | S p => n*n^p end
where "n ^ m" := (nat_power n m).

Fixpoint sum_f (f:nat->nat)(n:nat) : nat :=
  match n with 0 => f 0 | S p => f (S p)+sum_f f p end.

Lemma mult_0_inv : forall x y, x*y=0->x=0\coqor{}y=0.
intros x; case x; simpl; auto.
intros x' y; case y; simpl; auto; intros; discriminate.
Qed.

Lemma power_0_inv : forall x n, x ^ n = 0 -> x = 0.
intros x n; elim n.
simpl; intros; discriminate.
intros n' IH H; case (mult_0_inv x (x ^ n') H); subst; auto.
Qed.

Lemma sum_of_powers1 : forall x n, 
     x*sum_f (nat_power x) n + 1 = x^(n+1)+sum_f (nat_power x) n.
intros x n; elim n; simpl; repeat rewrite mult_1_r; auto.
simpl; intros p IH; rewrite mult_plus_distr_l; rewrite <- plus_assoc;
rewrite IH; rewrite <- (plus_comm 1); simpl; ring.
Qed.

Lemma sum_of_powers :  forall x n, 1 <= x ->
  (x-1)*sum_f (nat_power x) n = x^(n+1)-1.
intros x n Hx; apply plus_reg_l with (1*sum_f(nat_power x) n).
rewrite <- mult_plus_distr_r; rewrite <- le_plus_minus; auto.
apply plus_reg_l with 1.
assert (H' : 1 +(1*sum_f (nat_power x) n+ (x^(n+1)-1))=
             1*sum_f(nat_power x) n + (1+(x^(n+1)-1))).
ring.
rewrite H'; clear H'; rewrite <- le_plus_minus.
rewrite (plus_comm 1); rewrite sum_of_powers1; ring.
case (zerop (x^(n+1))); auto.
intros H; assert (x = 0). apply (power_0_inv _ _ H). omega.
Qed.

Require Import ZArith.  Open Scope Z_scope.

Fixpoint Zsum_f (f:nat->Z)(n:nat) : Z :=
  match n with 0%nat => f 0%nat | S p => f (S p)+Zsum_f f p end. 

Theorem Zsum_of_powers :  forall x n,
 (x-1)*Zsum_f (fun n =>Zpower x (Z_of_nat n)) n = x^(Z_of_nat n+1)-1.
intros x n; elim n.
unfold Zsum_f, Z_of_nat; simpl (0+1); ring.
intros p IHp.
assert (Hinz : (Z_of_nat (S p))=(1+Z_of_nat p)).
rewrite inj_S; unfold Zsucc; ring.
unfold Zsum_f; fold Zsum_f; rewrite Hinz; repeat rewrite Zpower_exp;
try (rewrite Zmult_plus_distr_r; rewrite IHp); 
repeat rewrite Zpower_exp;  auto with zarith; ring.
Qed.

Require Import Reals. Open Scope R_scope.

Fixpoint Rsum_f (f:nat->R)(n:nat) : R :=
 match n with 0%nat => f 0%nat | S p => f (S p)+Rsum_f f p end.

Lemma Rsum_of_powers : forall x n, (x-1)*Rsum_f (pow x) n = x^(n+1)-1.
intros x n; elim n.
simpl; ring.
intros p IHp; simpl; rewrite Rmult_plus_distr_l; rewrite IHp.
rewrite pow_add; ring.
Qed.
\end{alltt}
}

\end{document}